\documentclass[11pt,twoside]{article}
\usepackage{asp2004}
\usepackage{psfig}
\usepackage{epsf}
\usepackage{graphics}
\usepackage{lscape}
\def\edcomment#1{\iffalse\marginpar{\raggedright\sl#1\/}\else\relax\fi}
\reversemarginpar

\newcommand {\hi} {H\,{\small I}}
\newcommand {\kms} {km s$^{-1}$}
\newcommand {\ci}{$^{\circ}$}

\begin{document}
\title{The extra-planar neutral gas in the edge-on spiral galaxy NGC\,891}

\author{Filippo Fraternali}
\affil{Theoretical Physics, University of Oxford (UK)}

\author{Tom Oosterloo}
\affil{ASTRON, Dwingeloo (NL)}

\author{Renzo Sancisi}
\affil{INAF-Bologna (I) \& Kapteyn Institute, Groningen (NL)}

\author{Rob Swaters}
\affil{Department of Astronomy, University of Maryland (MD)}

\begin{abstract}
We present neutral hydrogen observations of the nearby
edge-on spiral galaxy NGC\,891 which show extended extra-planar emission 
up to distances of 15 kpc from the plane.
3D modeling of the galaxy shows that this emission comes from 
halo gas rotating more slowly than the gas in the disk.
We derive the rotation curves of the gas above the plane and find a
gradient in rotation velocity of $-$15 \kms\ kpc$^{-1}$.
We also present preliminary results of a galactic fountain model 
applied to NGC\,891.

\end{abstract}
\thispagestyle{plain}

\section{Introduction}

NGC\,891 is one of the best known and studied nearby edge-on 
spiral galaxies. 
It is at the distance of 9.5 Mpc, is classified as 
a Sb/SBb, and it is often referred to as a galaxy 
very similar to the Milky Way \citep{vdk81}.
Because of its very high inclination \citep[i$\geq$88.6\ci, ][]{rup91}
it is very suitable for the study of the
distribution and kinematics of the gas above the plane.

NGC\,891 has been the subject of numerous
studies at different wavelengths that have led to the detection of
various halo components:
an extended radio halo \citep*{all78, hum91},
an extended layer of diffuse ionised gas (DIG) \citep[e.g.][]{det90}
and diffuse extra-planar hot gas \citep{bre94}.
Also ``cold'' ISM components have been detected in the 
halo such as \hi\ \citep*{swa97}, dust \citep{how99} and
CO \citep{gar92}.

Here we concentrate on the neutral gas and present results
from recent third generation 
\hi\ observations obtained with the Westerbork Synthesis Radio 
Telescope (WSRT).
NGC\,891 was first studied in \hi\ in the late seventies with
the WSRT and the presence of neutral gas seen in projection out of
the plane was reported \citep{san79}.
Subsequently, a new study with higher sensitivity 
showed that the extra-planar emission was very extended, up
to 5 kpc from the plane. 
3D modeling indicated that such emission was produced by a thick layer
of neutral gas rotating more slowly than the gas in the disk
\citep{swa97}. 

Since then, several other studies have confirmed the presence of neutral
gas in the halos of spiral galaxies.
It has been detected in edge-on or nearly edge-on systems
(e.g. UGC\,7321, \citet{mat03} and NGC\,253, \citet{boo04}), as well as
in galaxies viewed at different inclination angles 
\citep[e.g.\ NGC\,2403,][]{fra01}. 
Indications of vertical gradients in rotation velocity have been
found in several galaxies also in the ionised gas (e.g.\ NGC\,891,
\citet{pil94}, \citet{ran97}; NGC\,5055, \citet{ran00}).

In the first part of the paper we present 
the new \hi\ observations of NGC\,891 together with a 3D modeling 
of the \hi\ layer. 
In the second part (Section 3) we study 
the kinematics of the extra-planar gas and, in the third one (Section 4), 
we present results from a dynamical model of the extra-planar gas.

\section{HI observations}

\begin{figure}[!ht]
\plotone{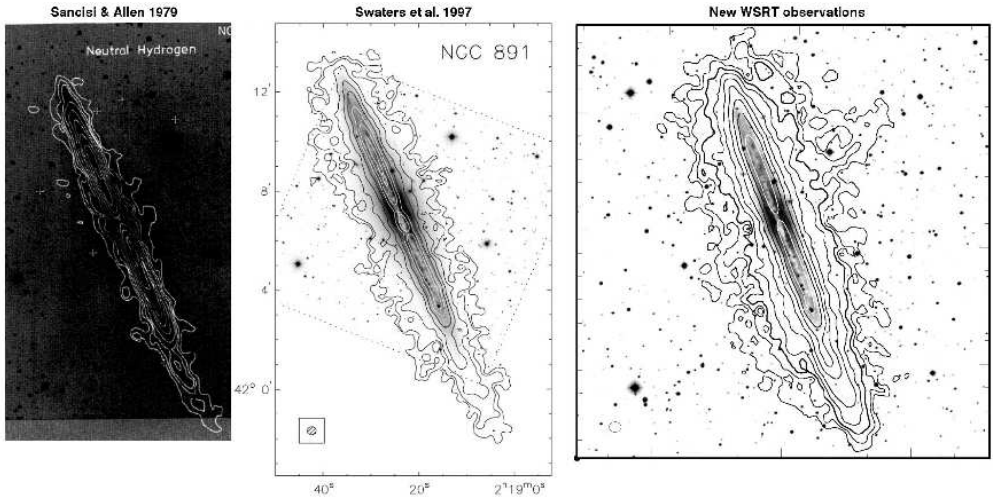}
\caption{
Total \hi\ map of NGC\,891 (right) obtained from our new WSRT observations
compared with the total \hi\ maps of the previous observations 
\citep{san79,swa97}.
Contours of the new total \hi\ map 
are: 1.7, 4.5, 9, 18.5, 37, 74, 148, 296.5, 593 $\times$
10$^{19}$ atoms cm$^{-2}$.
}
\end{figure}

\begin{figure}[!ht]
\plotone{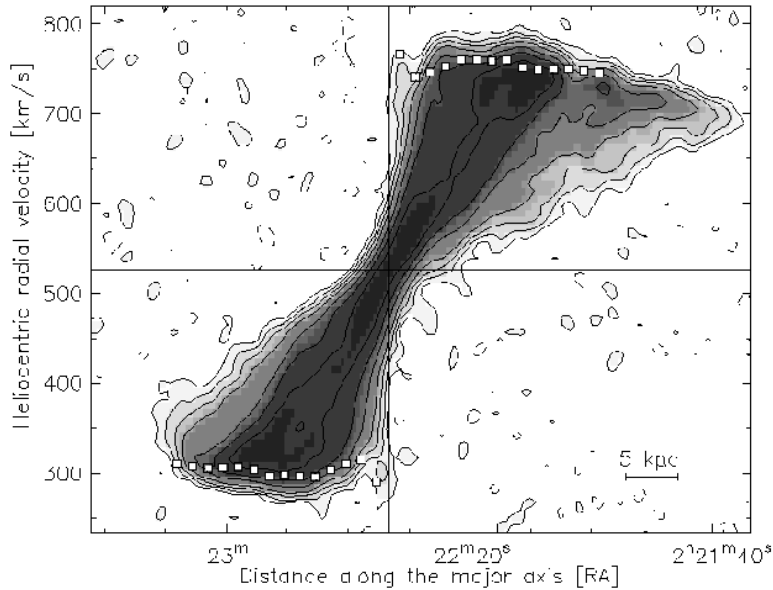}
\caption{ 
\hi\ position-velocity diagram along the major axis of NGC\,891. 
The white squares show the rotation curve derived using only the 
approaching side of the galaxy.
}
\end{figure}

The new observations of NGC\,891 have been carried out during
the second semester of 2002 with the
Westerbork Synthesis Radio Telescope (WSRT) and a total integration
time of about 200 hrs.
With this long integration time we reach a sensitivity 
(r.m.s.\ noise per channel = 0.22 mJy/beam at 28$'' \times 16.5$ \kms\
resolution) that is about a factor 4 better than the previous
observations of \citet{swa97}.
A complete presentation of the observations will be given in 
Oosterloo et al.\ (in preparation). 

Figure 1 (right panel) shows the new total \hi\ map of NGC\,891 at
28$''$ ($\sim$1.3 kpc). 
\hi\ emission is detected at a projected distance of as far as 15 kpc 
from the plane (see the spur in the N-W side of the disk).
The size of the disk itself (in the plane) is very similar to that
reported by \citet{swa97} and \citet{san79}
suggesting that we have possibly reached the edge of the 
\hi\ disk, especially on the N-E side.
The emission above the plane, instead, is significantly more extended
than in previous observations and almost everywhere
extends up to 10 kpc above and below the plane. 

Figure 2 shows the position-velocity plot along the major axis of NGC\,891
at 28$''$ resolution with the rotation curve (white squares) overlaid.
The rotation curve was derived with the method 
described in section 3 using only the N-E (approaching) side of the galaxy. 
The kinematics of the receding side within a radius of $\sim$6 kpc is
very similar to that of the approaching one.
At larger radii the velocity is apparently declining.
However, we cannot be sure that the gas in this extension is at the
line of nodes and, therefore, the derivation of a rotation curve in
that region is not possible. 
In the inner regions of the galaxy
(about 1-2 kpc) we confirm the presence of a fast 
rotation disk or ring \citep[see also][]{swa97}.

\subsection{The extra-planar gas}

\begin{figure}[!ht]
\plotone{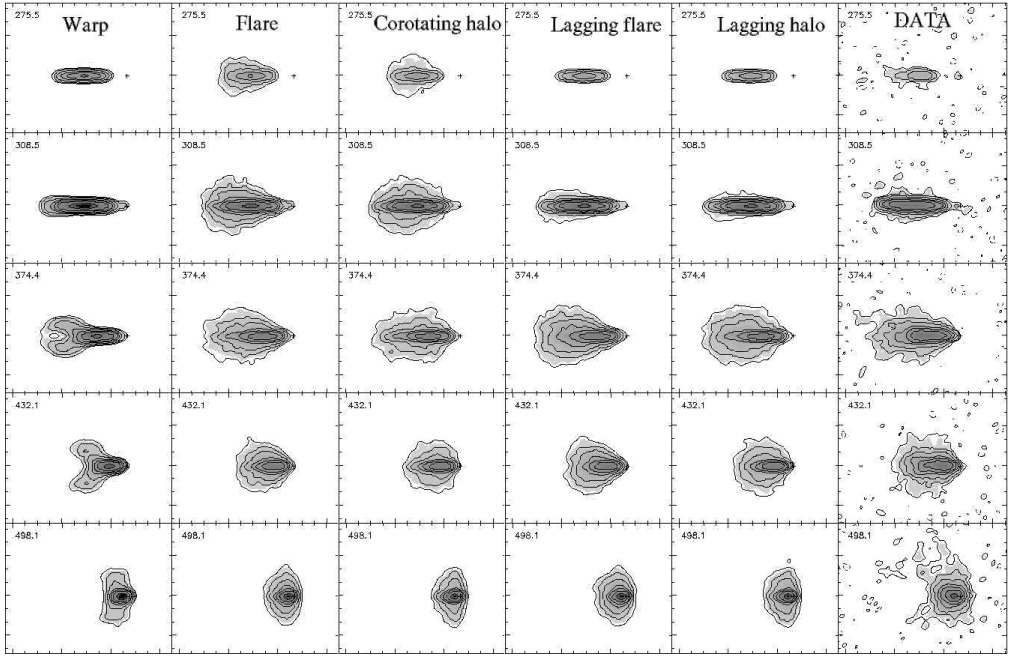}
\caption{
Channel maps for NGC\,891 compared with galaxy models (see text).
}
\end{figure}

The total \hi\ map of NGC\,891 (Figure 1) shows extended
emission in the direction perpendicular to the plane.
Is this emission coming from gas located in the halo of the galaxy 
or is it the result of projection effects?
Here we address this question using a 3D
modeling technique similar to that of \citet{swa97}.

In Figure 3 we compare some of the observed channels maps of NGC\,891 (right hand column) with models of the \hi\ layer.
From the left hand column they are:
1) a warp along the line of sight, i.e.\ a change of the inclination
   angle of the disk, from 90 to 60 degrees, in the outer parts;
2) a flaring (increasing thickness) of the outer disk from a FWHM of 0.5 
kpc up to $\sim$6 kpc;
3) a two-component model with thin disk + thick (FWHM$\sim$6 kpc) disk
   corotating;
4) and 5) two-component models with the thick disk rotating more
slowly (35 \kms) than the disk. 
The models in columns 4 and 5 differ only for the radial density
distribution of the thick component: one (5) is the same (scaled) as 
that of the gas in the disk and the other (4) has a depression in the
central regions.

Of the models reported here, only those in columns 4 and 5 give a
reasonable reproduction of the data. 
In particular the warp model does not reproduce the shape of channel
maps at 374 and 432 \kms, while the flare and corotating model do not
reproduce the thin channels of the two top rows (at 275.5 and 308.5
\kms).
This clearly indicates that the extra-planar emission in NGC\,891 is
produced by a thick (FWHM$\sim$6 kpc) layer of \hi\ rotating more
slowly than the gas in the plane.
The halo is possibly relatively denser than the disk in the outer
parts (with a radial density distribution somewhat in between the
models in column 4 and 5).

\section{Kinematics of the extra-planar gas}

In the previous section we have presented some simple galaxy models
showing that the emission above the plane in NGC\,891 comes from gas
which is located in the halo region and rotates more slowly (is
lagging) with respect to the gas in the plane.
Here we quantify this lag by deriving the 2D rotation velocity field 
(rotation surface) or rotation curves at different
heights from the plane.

\begin{figure}[!ht]
\plotone{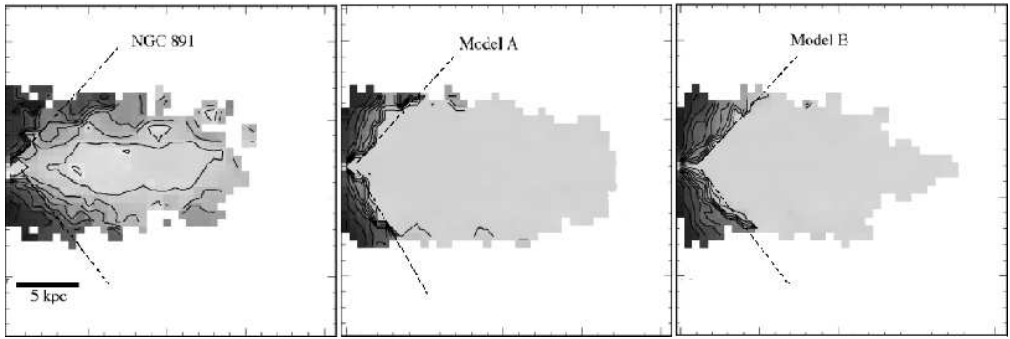}
\caption{Rotation velocity surfaces for NGC\,891 (N-E side) and two
model galaxies obtained with the envelop-tracing technique.
The centre of the galaxy is at the middle of the left edge of each
figure. 
The models have constant rotation velocity in R and z. 
In the inner regions (outlined by the dashed lines) the envelop-tracing
fit does not work properly (i.e.\ it does not return the input constant
rotation velocity).
}
\end{figure}

A rotation curve of a spiral galaxy seen edge-on 
is usually derived using an envelop-tracing method 
i.e.\ by taking the highest measured
velocity along the line of sight as the rotation velocity
\citep[e.g.][]{san79}.
This method was, for instance, used to derive the rotation curve shown
in Figure 2.
In an ideal situation (high S/N, galaxy perfectly edge-on, pure
rotation, etc.) the highest velocity measured along the
line of sight is indeed produced by the gas at the line of nodes
(i.e.\ rotation velocity). 
However, the real situation can be very different.
In particular it is possible that the emission from the gas at
the line of nodes is below detection and, as a consequence,
the rotation velocity is underestimated.
This could easily happen when observing the gas above the plane where
the density is lower and especially in cases of central \hi\
depressions or holes.

Figure 4 (left panel) 
shows the rotation velocity surface (contours with constant $v_\phi$)
of NGC\,891 derived with the 
envelop-tracing method for the N-E side of the galaxy.
Can this rotation velocity field be fully trusted?

In order to test this, we have constructed model galaxies with known 
(flat) rotation velocity in R and z.
Figure 4 (central and right panels) 
shows two rotation velocity surfaces obtained from two of these
models.
These surfaces are obtained using exactly the same fit and fitting 
parameters as for the one of NGC\,891 (left panel).
The two models have different radial density distributions for the 
extraplanar gas. 
These were obtained from the data from above (N-W side, 
Model A) and below (S-E side, Model B) the disk.
The results show that the fit works very well in a large
part of the galaxy giving a constant value of the velocity in R and z 
(uniform lowest gray level).
The fit does not return the input constant rotation velocity 
with z only in the inner regions outlined by the dashed lines in
Figure 4.
There is very little difference between the results 
obtained with the two different radial density
distributions.
These results were used to exclude the 
inner regions in the 2D rotation velocity field of NGC\,891 (left
panel).  

\begin{figure}[!ht]
\plotone{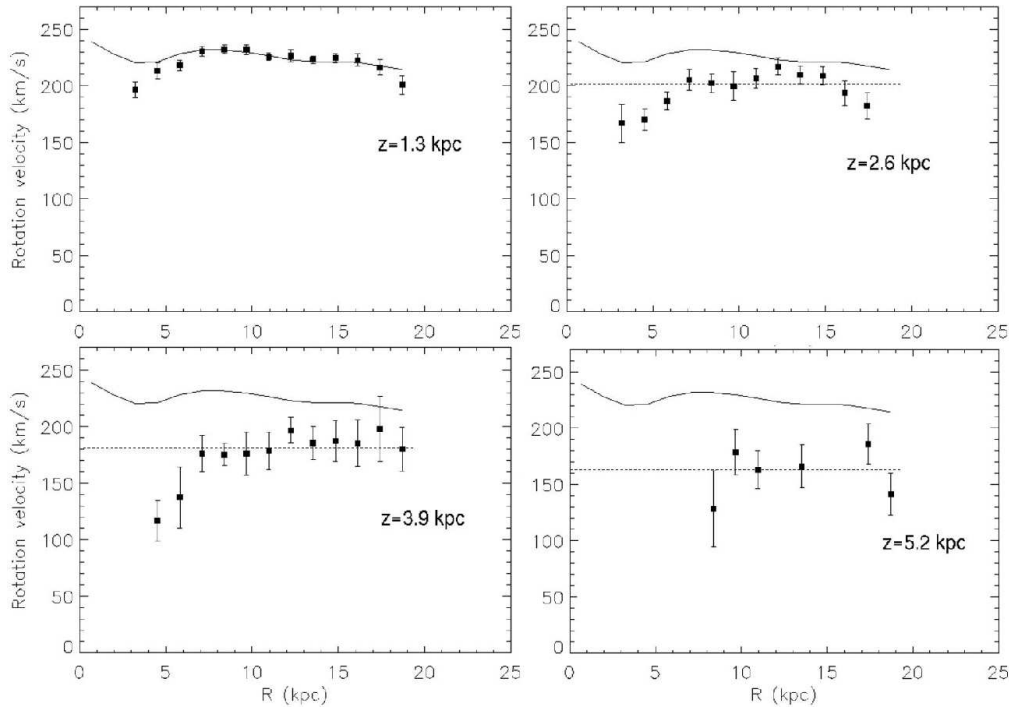}
\caption{Rotation curves for NGC\,891 at various distances from the
plane. The solid line shows the rotation curve in the plane 
and the dashed lines outline the value of the flat part of the
rotation curves.
Only the N-E side of the galaxy has been used to derive these
curves.
}
\end{figure}

Figure 5 shows the results for NGC\,891 plotted as rotation curves at various
distances from the plane after the uncertain inner points
have been removed. 
The solid line shows the (smoothed) rotation curve in the plane while
the dashed lines outline the value of the flat part of the rotation 
curves at the different heights.
The halo of NGC\,891 appears to corotate up to 1.3 kpc, then
it starts to lag with respect to the disk, the lagging increasing with
height from the plane. 
Given the limited angular resolution, the apparent corotation below
1.3 kpc may be the effect of beam-smearing.
The gradient in rotation velocity is roughly 15 \kms\ kpc$^{-1}$.
The West and East side of the galaxy do not show significant
differences.
A detailed description of this procedure and results will be given in
a forthcoming paper (Fraternali, in preparation).

\section{Dynamical models}

The presence of neutral gas in the halo of spiral galaxies is still
unexplained.
It can be either the result of a galactic fountain 
\citep{sha76}
and/or of accretion from the intergalactic medium \citep[e.g.\
][]{oor70}.
The processes involved are probably non-stationary and require
a dynamical modeling of the medium above the plane 
(see \citet{bar04} for a stationary model).
As a first approach to the problem we have considered a ballistic 
(particle-based) 
model for the gas expelled from the plane as a result of
star formation activity \citep[see also][]{col02}.
We have considered a multicomponent potential with stellar and gaseous
exponential disks, a r$^{1/4}$ bulge and a double power-law dark
matter halo with adjustable flatness \citep[in analogy to the galactic
models of][]{deh98}.

\begin{figure}[!ht]
\plotone{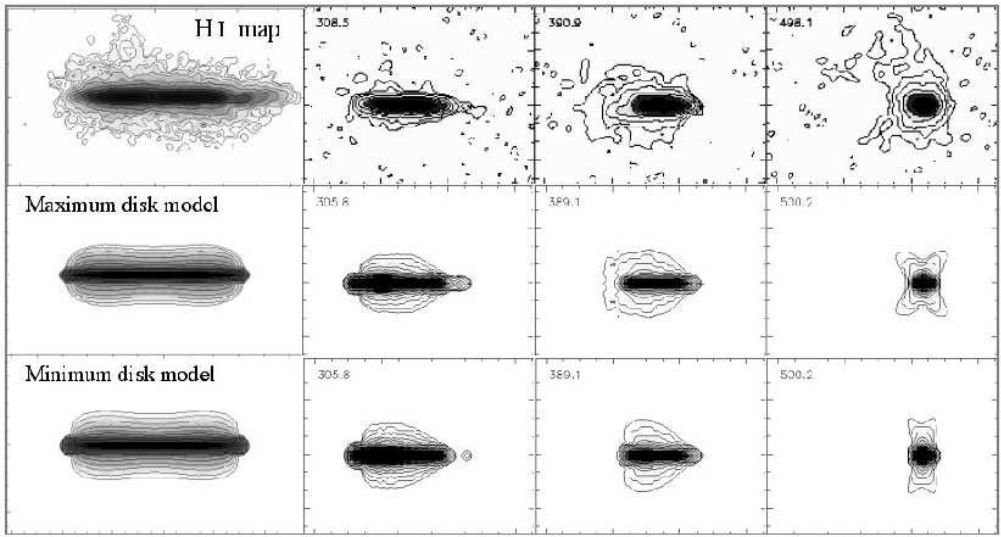}
\caption{Total \hi\ map and representative channel maps for NGC\,891
(top) compared to two ballistic models of galactic fountains.
The number in the upper left corners are heliocentric velocity
(\kms). 
}
\end{figure}

A C++ code was written to integrate the orbits of the particles and
project them along the line of sight. 
The particles are shot vertically with different kick velocities from
0 to a maximum v$_{max}$ that can depend on R.
For each $dt$ (time interval) the position and velocity of the particle
is projected along the line of sight and the integration is carried
out until the particle falls back to the plane.
The output of the code is a model cube that can be compared
with the data cubes. 
These models can be applied to external galaxies viewed at any
inclination angle as well as to the Milky Way (internal projection). 
Here we show a preliminary application to NGC\,891. 
Details and
improvements of the model to take into account also non-ballistic
effects and gas accretion from the intergalactic medium will be
presented in a forthcoming paper (Fraternali \& Binney, in
preparation).

Figure 6 shows the total \hi\ map and some representative channel maps
for NGC\,891, compared with the outputs of two ballistic models. 
The two models are for a maximum and a minimum disk fit of the rotation 
curve with B band M/L ratios of 7 and 1 respectively.
The parameters of the models have been ``tuned'' to reproduce as well
as possible the vertical distribution of the gas and the shape of the
channel maps.
This figure shows that both ballistic models 
can qualitatively reproduce the data.  
However, there seem to be problems in reproducing the thinness of the
channel maps around 300 \kms\, i.e.\ there is not enough lagging in
the halo.
This is not the effect of the potential considering that little
differences are visible when considering extreme choices of the
potential shape. 
This preliminary analysis suggests that the pure ballistic galactic 
fountain models fail to explain the amount of lagging in galactic 
halos and that other phenomena such as gas accretion may be important.

\section{Conclusions}

New \hi\ observations of the edge-on spiral galaxy NGC\,891 show the
presence of extraplanar neutral gas up to distances of 15 kpc from
the plane.
We have used these data to derive, for the first time, a 2D
rotation velocity field (in R and z) of an edge-on spiral galaxy.
We find that the extra-planar gas
is corotating with the gas in the disk up to about 1.3
kpc; beyond that it rotates more slowly by about 15 \kms\ kpc$^{-1}$.
Dynamical models of NGC\,891 show that a pure ballistic galactic
fountain can qualitatively reproduce the data. 
However, problems with missing low angular momentum material suggest
that other mechanisms such as accretion may also play a role.

\end{document}